\documentclass[graybox]{svmult}

\usepackage{type1cm}        
\usepackage{makeidx}         
\usepackage{graphicx}        
\usepackage{multicol}        
\usepackage[bottom]{footmisc}

\usepackage{newtxtext}       %
\usepackage{newtxmath}       

\makeindex             
                       
\usepackage{mathtools,pifont}
\usepackage{graphicx,color,dsfont,multirow}
\usepackage{tabularx,sectsty,pifont,booktabs,threeparttable}
\usepackage{bigdelim}   
\usepackage[caption=false]{subfig}                    

\usepackage[style=nature,citestyle=authoryear,sortcites=false,uniquename=init,backend=bibtex,natbib=true,maxbibnames=99,maxcitenames=2,language=american,giveninits=true]{biblatex}
\DefineBibliographyStrings{american}{phdthesis = {PhD dissertation}}

\newcommand{\cmark}{\ding{51}}

\newcolumntype{L}[1]{>{\raggedright\arraybackslash}p{#1}}
\newcolumntype{C}[1]{>{\centering\arraybackslash}p{#1}}
\newcolumntype{R}[1]{>{\raggedleft\arraybackslash}p{#1}}                

\begin{document}

\title*{Multiple Imputation for Nonignorable Item Nonresponse in Complex Surveys Using Auxiliary Margins}
\titlerunning{Multiple Imputation for Nonignorable Item Nonresponse in Complex Surveys}
\author{Olanrewaju Akande and Jerome P. Reiter}
\institute{Olanrewaju Akande \at Social Science Research Institute, Duke University, Durham, NC, \email{olanrewaju.akande@duke.edu}
\and Jerome P. Reiter \at Department of Statistical Science, Duke University, Durham NC, \email{jreiter@duke.edu}}
%
%
\maketitle

\abstract{We outline a framework for multiple imputation of nonignorable item nonresponse when the marginal distributions of some of the variables with missing values are known. In particular, our framework ensures that (i) the completed datasets result in design-based estimates of totals that are plausible, given the margins, and (ii) the completed datasets maintain associations across variables as posited in the imputation models.  To do so, we propose an additive nonignorable model for nonresponse, coupled with a rejection sampling step.  The rejection sampling step favors completed datasets that result in design-based estimates that are plausible given the known margins.  We illustrate the framework using simulations with stratified sampling.}


\section{Introduction}
Many surveys suffer from item nonresponse that may be nonignorable.  This can complicate analysis or dissemination of survey data.  In some settings, we can leverage auxiliary information from other data sources to help adjust for the effects of nonignorable nonresponse. 
For example, suppose that in a simple random sample, a question on sex  suffers from item nonresponse, so that 70\% of the respondents are women.  Suppose we know that the target population includes 50\% men and 50\% women. This implies that respondents with missing values of sex are more likely to be men than women.  Thus, if we impute values for the missing sexes, we should impute more ``male'' than ``female''.      

Generalizing this example, we desire to leverage reliable estimates of low-dimensional margins for variables with item nonresponse---available, for example, from high quality surveys or administrative databases---when imputing missing items.  However, we do not want to use solely these population margins to inform the imputations.  We also should take advantage of observed information in other variables, so as to preserve multivariate relationships as best as possible.  In the case where the data are from a complex survey, we also need to somehow account for the survey design weights in the imputations \citep{reitragukin, ZhouEtAl2016}. We are not aware of any principled ways to do all this simultaneously when performing multiple imputation for item nonresponse.

In this chapter, we propose a framework for multiple imputation of missing items in complex surveys that leverages auxiliary margins.  Our approach is to use the auxiliary margins to identify additive nonignorable (AN) models \citep{HiranoEtAl1998, HiranoEtAl2001}, with an additional requirement that the completed datasets result in plausible design-based estimates of the known margins.  We do so by fusing the AN model with large sample results under frequentist (survey-weighted) paradigms. In this way, we ensure that imputations are influenced by relationships in the data and the auxiliary information, while being faithful to the survey design through survey weights. 

Our work connects to several areas of research in which Stephen Fienberg made key contributions. In particular, the methods are examples of using marginal information \citep{fienberg70, chenfienberg74}, and of course handling missing values \citep{bishopfienberg69, fienberg72, chenfienberg76}, in the analysis of contingency tables. Our approach also uses Bayesian techniques for official statistics and survey sampling, a perspective that he championed for many areas including disclosure limitation, record linkage, and the analysis of categorical data.

The remainder of this chapter is organized as follows. In Section \ref{ANmodel}, we review the AN model. In Section \ref{framework}, we present our approach. In Section \ref{simulations}, we illustrate the performance of the approach using simulation studies with stratified sampling. In Section \ref{discussion}, we conclude and discuss possible extensions.  For clarity, we present the methodology for data that does not have unit nonrespondents.  We discuss extensions to scenarios including unit nonrespondents in Section \ref{discussion}.

\section{Review of the AN Model} \label{ANmodel}

Our review the AN model closely follows the review in  \citet[Chapter 4]{Akande2019}.   For additional discussion of the AN model, see \citet{nevo2003using, bhattacharya2008inference, DasEtAl2011, dengreiter, schifelingcheng, si:reiter:hillygus15, SadinleReiter2019}. 
Although the AN model was developed originally for handling nonignorable attrition in longitudinal studies with refreshment samples \citep{HiranoEtAl1998, dengreiter}, it can be applied to our setting by viewing the data from the refreshment samples as auxiliary information, as we now describe.

\subsection{Notation} \label{ANmodel:notation}
Let $\cal{D}$ comprise data from the survey of $i = 1, \ldots, n$ individuals, and $\cal{A}$ comprise data from the auxiliary database. Let $X = (X_1, \ldots, X_p)$ represent the $p$ variables in both $\cal{A}$ and $\cal{D}$, where each $X_k = (X_{1k},\ldots,X_{nk})^T$ for $k = 1, \ldots, p$.
Let $Y = (Y_1, \ldots, Y_q)$ represent the $q$ variables in $\cal{D}$ but not in $\cal{A}$, where each $Y_k = (Y_{1k},\ldots,Y_{nk})^T$ for $k = 1, \ldots, q$.
We assume that $\cal{A}$ only contains sets of marginal distributions for variables in $X$, summarized from some external database and measured with negligible error. 
We disregard variables in $\cal{A}$ but not $\cal{D}$, as the margins for these variables generally do not provide much information about the missing values in $\cal{D}$.

We also introduce variables to account for item nonresponse. For each $k = 1, \ldots, p$, let $R^x_k = (R^x_{1k},\ldots,R^x_{nk})^T$, where each $R^x_{ik} = 1$ if individual $i$ would not respond to the question on $X_k$ in $\cal{D}$, and $R^x_{ik} = 0$ otherwise. Similarly, for each $k = 1, \ldots, q$, let $R^y_k = (R^y_{1k},\ldots,R^y_{nk})^T$, where each $R^y_{ik} = 1$ if individual $i$ would not respond to the question on $Y_k$ in  $\cal{D}$ and $R^y_{ik} = 0$ otherwise. 

Finally, for simplicity, we use generic notations such as $f$ and $\eta$ for technically different functions and parameters respectively, although their actual meanings should be clear within each context. For example, $f$, $\eta_{0}$, and $\eta_{1}$ need not be the same in the conditional probability mass functions $\Pr(X_1= 1|Y_1) = f(\eta_{0} + \eta_{1}Y_1)$ and $\Pr(Y_1 = 1 | X_1) = f(\eta_{0} + \eta_{1}X_1)$.

\subsection{AN model specification} \label{ANmodel:ModelSpecification}
To make the AN model specification easy to follow, we work with an example where $\cal{D}$ comprises only two binary variables, $X_1$ and $Y_1$. Following our notation, $\cal{A}$ contains the auxiliary marginal distribution for $X_1$ but no auxiliary marginal distribution for $Y_1$. For simplicity, we also suppose $X_1$ suffers from item nonresponse but $Y_1$ is fully observed. Thus, we need a model for $R^x_1$, the fully observed vector of item nonresponse indicators for $X_1$.  We assume that we do not need to include a model for $R^y_1$, since there is no nonresponse in $Y_1$. The observed and auxiliary data take the form shown in Table \ref{ANmodel:marg}(a).  The incomplete contingency table representing the joint distribution of $(X_1,Y_1,R^x_1)$, with observed and auxiliary marginal probabilities excluded, is shown in Table \ref{ANmodel:marg}(b). 

Due to the empty cells in the contingency table in Table \ref{ANmodel:marg}(b), we cannot fit a fully saturated model to these data. To see this, we use a pattern mixture model factorization \citep{GlynnEtAl1986,littlepatmix} to characterize the joint distribution of $(X_1,Y_1,R^x_1)$. The factorization, which we write as
\begin{equation} \label{ANmodel:patternmix}
	\begin{split}
		\Pr(X_1=x,Y_1=y,R^x_1=r) & = \Pr(X_1=x | Y_1=y,R^x_1=r) \\
		& \times \Pr(Y_1=y | R^x_1=r)\Pr(R^x_1=r),
	\end{split}
\end{equation}
can be fully parameterized using seven parameters: the four values of $\theta_{yr} = \Pr(X_1=1|Y_1=y,R^x_1=r)$, $\pi_{r} = \Pr(Y_1=1 | R^x_1=r)$ and $q = \Pr(R^x_1=1)$.
\begin{table}[t] 
	\centering
	\caption{Set-up for the AN model with two binary variables $Y_1$ and $X_1$. $Y_1$ is fully observed, and $X_1$ suffers from item nonresponse.  We know the population margin for $X_1$. Here, ``\cmark'' represents observed components and ``\textbf{?}'' represents missing components.} \label{ANmodel:marg}
	\subfloat[][Data]{ 
		\begin{tabular}{R{2.5cm}|C{0.6cm}|C{0.6cm}|C{0.6cm}|} 
			\cline{2-4}
			& $X_1$  & $Y_1$ & $R^x_1$ \\ \cline{2-4}
			\multirow{2}{*}{Original data} \ldelim\{{2}{2mm} & \cmark & \multirow{2}{*}{\cmark} & 0 \\ \cline{2-2} \cline{4-4}
			& \textbf{?} & & 1 \\\cline{2-4}
			Auxiliary margin $\rightarrow$ & \cmark & \textbf{?} & \textbf{?} \\ \cline{2-4}
	\end{tabular}} 
	\qquad 
	\subfloat[][Contigency table]{
		\begin{tabular}{|C{1cm}|C{1cm}|C{1cm}|C{1cm}|C{1cm}|} \hline
			& \multicolumn{2}{c|}{$R^x_1=0$} & \multicolumn{2}{c|}{$R^x_1=1$} \\ \cline{2-5}
			& $X_1=0$ & $X_1=1$ & $X_1=0$ & $X_1=1$ \\ \cline{1-5}
			$Y_1=0$ & \cmark & \cmark & \textbf{?} & \textbf{?} \\ \cline{1-5}
			$Y_1=1$ & \cmark & \cmark  & \textbf{?} & \textbf{?} \\ \cline{1-5}
	\end{tabular}}
\end{table}
Five of the seven parameters, that is, $q$, $\pi_0$, $\pi_1$, $\theta_{00}$, and $\theta_{10}$, can be directly estimated from the observed data alone, as long as the sample data is representative of the target population. Unfortunately, the observed data contain no information about $\theta_{01}$ and $\theta_{11}$.  We need to make assumptions about the missingness mechanism to estimate the full joint distribution. 
For example, we could set  $\theta_{01} = \theta_{00}$ and $\theta_{11} = \theta_{10}$, resulting in a missing at random (MAR) mechanism.  

The auxiliary marginal distributions provide information that we can use to specify such identifying assumptions. In our two-variable example, the auxiliary marginal distribution of $X_1$ provides one linear constraint about $\theta_{01}$ and $\theta_{11}$. 
We write this constraint as
\begin{equation} \label{ANmodel:linearconst}
	\Pr(X_1=1) - \Pr(X_1=1,Y_1=y,R^x_1=0) = q \left[ \theta_{01}(1-\pi_{1}) + \theta_{11}\pi_{1}\right].
\end{equation}
Although \eqref{ANmodel:linearconst} does not provide enough information to identify both $\theta_{01}$ and $\theta_{11}$, it does increase the number of estimable parameters from five to six. 

The AN model takes advantage of this additional constraint. In particular, the AN model assumes that the reason for item nonresponse in $X_1$ depends on $X_1$ and $Y_1$ through a function that is additive in $X_1$ and $Y_1$. We have 
\begin{align}
	\phantom{\Pr(R^x_1 = 1 | X_1,Y_1) =}
	&\begin{aligned} \label{ANmodel:selection01}
		\mathllap{(X_1,Y_1) \sim} & \ f(X_1,Y_1 | \Theta)
	\end{aligned}\\
	&\begin{aligned} \label{ANmodel:selection02}
		\mathllap{\Pr(R^x_1 = 1 | X_1,Y_1) =} & \ h(\eta_{0} + \eta_{1}X_1 + \eta_{2} Y_1),
	\end{aligned}
\end{align}
where $\Theta$, $\eta_0$, $\eta_1$, and $\eta_2$ represent the parameters in $f$ and $h$. Here, $h(a)$ should be a strictly increasing function satisfying $\lim_{a \rightarrow -\infty}h(a) = 0$ and $\lim_{a \rightarrow \infty}h(a) = 1$. The models in (\ref{ANmodel:selection01}) and (\ref{ANmodel:selection02}) represent a selection model factorization \citep{Little1995} of the joint distribution of $(Y_1,X_1,R^x_1)$, instead of the pattern mixture factorization in (\ref{ANmodel:patternmix}). \citet{HiranoEtAl2001} prove that the AN model is likelihood-identified for general distributions, such as  probit and logistic regression models. The interaction term between $X_1$ and $Y_1$ is not allowed, as additivity is necessary to enable identification of the model parameters. 

The AN model is appealing in that it includes ignorable and nonignorable models as special cases. For example, $(\eta_{1} = 0, \eta_{2} = 0)$ results in a missing completely at random (MCAR) mechanism, $(\eta_{1} \neq 0, \eta_{2} = 0)$ results in a MAR mechanism, and $\eta_{2} \neq 0$ results in a missing not at random (MNAR) mechanism. In particular, $(\eta_{1} = 0, \eta_{2} \neq 0)$ results in the nonignorable model of \citet{HausmanWise1979}. This allows the data determine an appropriate mechanism from among these possibilities. The AN model does rely on the assumption of additivity of the response model in $X_1$ and $Y_1$, which may be reasonable in practice. \citet{dengreiter} describe sensitivity analysis for non-zero interaction effects.  \citet{HiranoEtAl2001} suggest results are not overly sensitive to the choice of $h$. 

It is possible to use mechanisms other than the AN model to estimate up to six unique parameters (in our example here). For example, one can set either $\theta_{01}$ or $\theta_{11}$ equal to zero. Setting $\theta_{01}=0$ but $\theta_{11}\neq 0$ implies that all nonrespondents cannot have $X_1=1$ whenever $Y_1=0$. On the other hand, setting $\theta_{11}=0$ but $\theta_{01}\neq 0$  implies that all nonrespondents cannot have  $X_1=1$ whenever $Y_1=1$. Both assumptions seem more restrictive than setting an interaction effect in the model for $R^x_1$ to zero, and we do not recommend adopting them unless the specific application at hand justifies such strong assumptions. As another example, one can set $\theta_{01} = \theta_{11} + b$ for some constant $b$.  With $b=0$, this equates to $\theta^\star = \theta_{01} = \theta_{11}$, which then simplifies (\ref{ANmodel:linearconst}) to
\begin{equation}
	\theta^\star = \dfrac{\Pr(X_1=1) - (1-q)\left[\theta_{00}(1-\pi_{0})+ \theta_{10}\pi_{0}\right]}{q}.
\end{equation}
This constraint  implies conditional independence between $Y_1$ and $X_1$ for nonrespondents.  This seems a strong assumption in general. 

These two options, as well as other mechanisms which we do not cover here, are seemingly more restrictive than the AN assumptions or do not maximize all available information. The AN model does not force analysts to make as many untestable assumptions as most of the other mechanisms do, while allowing analysts to estimate as many parameters as possible with auxiliary data. However, the AN model as developed by \citet{HiranoEtAl1998, HiranoEtAl2001} does not incorporate complex survey designs directly. We now extend the model to do so.

\section{Extending the AN Model to Account for Complex Surveys} \label{framework}

Let $N$ represent the number of units in the population from which the $n$ survey units in $\cal{D}$ are sampled. Let $W = (w_1, \ldots, w_n)$, where each $w_i$ is the base weight for the $i$th unit in the sample $\cal{D}$. Here, we let $w_i=1/\pi_i$, where $\pi_i$ is the probability of selection of the $i$th unit.  We present methods where weights are not subject to calibration or nonresponse adjustments, although one could use the approach for adjusted weights as well. Let the superscript ``$pop$'' represent the population counterparts of the survey variables. For example, $X^{pop}$ and $Y^{pop}$ represent the population based counterparts of $X$ and $Y$ respectively, where each $X_i \in X^{pop}$ and $Y_i \in Y^{pop}$. We do not observe values of $X^{pop}$ or $Y^{pop}$ for all non-sampled units in the population. 

To present the methodology, we continue to work with the two variable example in Section \ref{ANmodel:ModelSpecification}, with one minor modification. We now let $Y_1$ be a categorical variable with three levels, that is, $Y_1 \in \{1,2,3\}$. We do so to show that our approach can extend to non-binary variables. The data and incomplete contingency table take similar forms to Table \ref{ANmodel:marg}, with weights now included and $Y_1$ having three levels.

Following our discussions in Section \ref{ANmodel:ModelSpecification}, we once again cannot fit a fully saturated model to the data. However, we can uniquely estimate seven of the nine parameters in a fully saturated model.
Without any auxiliary information, we can fit the following model to the observed data as a default option within the missing data literature. We have
\begin{align}
	\phantom{\Pr(R^x_1 = 1 | X_1,Y_1) =}
	&\begin{aligned} \label{framework:selection1}
		\mathllap{Y_1 \sim} & \ f(\theta)
	\end{aligned}\\
	&\begin{aligned} \label{framework:selection2}
		\mathllap{\Pr(X_1 = 1 | Y_1) =} & \ g(\alpha_0 + \alpha_{1j} \mathds{1}[Y_{1}=j])
	\end{aligned}\\
	&\begin{aligned} \label{framework:selection3}
		\mathllap{\Pr(R^x_1 = 1 | X_1,Y_1) =} & \ h(\gamma_0 + \gamma_{1j} \mathds{1}[Y_{1}=j]),
	\end{aligned}
\end{align} 
resulting in a MAR mechanism, where $j = 1, 2,3$. We set $\alpha_{11} = 0$ and $\gamma_{11} = 0$ to ensure the model is identifiable; the model then only contains seven parameters as desired. For more flexibility however, we seek to fit a nonignorable model that includes $\gamma_2 X_{1}$ in (\ref{framework:selection3}), so that (\ref{framework:selection3}) becomes the AN model
\begin{align}
	\phantom{\Pr(R^x_1 = 1 | X_1,Y_1) =}
	&\begin{aligned} \label{framework:selection4}
		\mathllap{\Pr(R^x_1 = 1 | X_1,Y_1) =} & \ h(\gamma_0 + \gamma_{1j} \mathds{1}[Y_{1}=j] + \gamma_2 X_{1}).
	\end{aligned}
\end{align}
To do so, we need to incorporate at least one constraint on the remaining parameters. When the survey design is complex, it may not be sufficient to use the auxiliary margin to force an extra constraint on the remaining parameters as we did in Section \ref{ANmodel:ModelSpecification}, since that approach does not incorporate the survey weights directly. To account for the survey weights, we take a different approach.

In practice, the most common marginal information is the population total (or mean) of some of the variables. For example, for totals, we know that
\begin{equation} \label{populationtotal}
	T_X = \sum_{i=1}^{N} X_{i1}^{pop} = N \times \Pr(X_1^{pop} = 1),
\end{equation}
where $\Pr(X_1^{pop} = 1)$ is the true auxiliary marginal probability.
A classical design-unbiased estimator of $T_X$ in this case is the Horvitz-Thompson estimator \citep{HorvitzThompson1952}, henceforth referred to as HT estimator, which is
\begin{equation} \label{HTestimator}
	\hat{T}_X = \sum_{i \in \cal{D}} \dfrac{X_{i1}}{\pi_i} = \sum_{i \in \cal{D}} w_i X_{i1}.
\end{equation}
In large enough samples, finite population central limit theorems ensure that $\hat{T}_X$ is approximately normally distributed around $T_X$, with a variance $V_X$ that is estimated using design-based principles \citep{Fuller2009:Ch1}.  Thus, for fully observed data, we have 
\begin{equation} \label{constraint:obs}
	\sum_{i \in \cal{D}} w_iX_{i1} \sim N(T_X, V_X).
\end{equation}

When the data contain nonresponse, we cannot compute $\hat{T}_X$ directly.  However, it is reasonable to expect this unobserved value of $\hat{T}_X$ to be distributed around $T_X$ as governed by \eqref{constraint:obs}.  Thus, when we impute the missing values for $X_1$, it is reasonable to require any completed dataset to produce a value of $\hat{T}_X$ that is plausible under \eqref{constraint:obs} as well.  We operationalize this logic as follows.  
For all $i \in \cal{D}$, let $X_{i1}^\star = X_{i1}$ when $R^x_{i1} = 0$, and let $X_{i1}^\star$ be an imputed value when $R^x_{i1} = 1$. We impose the probabilistic constraint,
\begin{equation} \label{constraint}
	\sum_{i \in \cal{D}} w_iX_{i1}^\star \sim N(T_X, V_X).
\end{equation}
In this way, we favor imputations consistent with (\ref{constraint}) when generating imputed values for $X$ under the posterior predictive distribution implied by (\ref{framework:selection1}), (\ref{framework:selection2}) and (\ref{framework:selection4}). 
Using a probabilistic constraint, as opposed to a deterministic constraint that $\hat{T}_X$ be as close to $T_X$ as possible, reflects uncertainty about $\hat{T}_X$ more appropriately.  
Here, we assume $V_X$ is pre-specified and treated as known; for example, it could be based on previous knowledge or an average of estimates from preliminary sets of completed data.  We discuss considerations with unknown $V_X$ further in Section \ref{discussion}. 

We incorporate  (\ref{constraint}) into a Markov chain Monte Carlo (MCMC) sampler for the model parameters through a Metropolis algorithm. At each MCMC iteration $t$, let the current draw of each $X_{i1}^\star$ be $X_{i1}^{\star(t)}$ and let $\hat{T}_X^{\star(t)} = \sum_{i \in \cal{D}} w_iX_{i1}^{\star(t)}$. We use the following sampler at iteration $t+1$.
\begin{enumerate}
	\item[S1.] For all $i \in \cal{D}$, i.e., $i = 1, \ldots, n$, set $X_{i1}^\star = X_{i1}$ when $R^x_{i1} = 0$. When $R^x_{i1} = 1$, generate a candidate $X_{i1}^\star$ for the missing $X_{i1}$ from the following posterior predictive distribution implied by (\ref{framework:selection2}) and (\ref{framework:selection4}). We have 
	\begin{equation}
		\Pr(X_{i1}^\star = 1 | \ldots) \propto g(\alpha_0 + \alpha_{1j} \mathds{1}[Y_{i1}=j]) \ h(\gamma_0 + \gamma_{1j} \mathds{1}[Y_{i1}=j] + \gamma_2 X_{i1}^\star),
	\end{equation}
	    using the current posterior draws of the parameters at iteration $t+1$, where ``\ldots'' represents conditioning on all other variables and posterior draws of all parameters in the model.
	\item[S2.] Let $\hat{T}_X^\star = \sum_{i \in \cal{D}} w_iX_{i1}^\star$. Calculate the acceptance ratio, 
	\begin{equation}
		p =  \dfrac{N(\hat{T}_X^\star; T_X, V_X)}{N(\hat{T}_X^{\star(t)}; T_X, V_X)}.
	\end{equation}
	\item[S3.] Draw a value $u$ from $u \sim Unif(0,1)$. If $u \leq p$, accept the proposed candidate $(X_{i1}^\star, \ldots, X_{in}^\star)$, and set $X_{i1}^{\star(t+1)} = X_{i1}^\star$ for $i = 1,\ldots, n$. Otherwise, reject the proposed candidate, and set $X_{i1}^{\star(t+1)} = X_{i1}^{\star(t)}$ for $i = 1,\ldots, n$.
\end{enumerate}	
Intuitively, these steps reject completed datasets that yield highly improbable design-based estimates of $T_X$, while simultaneously allowing us to estimate $\gamma_2 X_{1}$ in (\ref{framework:selection4}). Although (\ref{constraint}) provides a stochastic constraint, whereas using the auxiliary margins as in Section \ref{ANmodel:ModelSpecification} forms linear constraints, $\gamma_2 X_{1}$ is still estimable when using (\ref{constraint}), as we show using the simulations in Section \ref{simulations}.

We recommend that analysts monitor the acceptance ratio of the missing data sampler in Steps S1 to S3, as with any Metropolis sampler. In cases where the acceptance ratio is considerably low, analysts can inflate or tune $V_X$ or consider other methods of generating more realistic imputations from the implied posterior predictive distribution. In our simulation scenarios in Section \ref{simulations}, there is no need to do so as the samplers mix adequately. We do not worry about cases where the acceptance ratio is high because we view (\ref{constraint}) as a constraint rather than a target distribution.
Therefore, we interpret a high acceptance ratio as the sampler doing a good job of generating imputations that respect the survey design, as desired.

\section{Simulations with Stratified Sampling} \label{simulations}

In this section, we illustrate the approach described in Section \ref{framework} via simulation studies with stratified sampling. 
We create ten populations, each of size $N=50000$ split into two strata: 70\% of units are in stratum 1 ($N_1 = 35000$), and 30\% of units  are in stratum 2 ($N_2 = 15000$). For each observation in each population, we generate values of a three-valued $Y_1$ and binary $X_1$ using  
\begin{align}
	\phantom{\Pr[X_{i1} | Y_{i1}] \sim} 
	&\begin{aligned} \label{simulations:probit1}
		\mathllap{Y_{i1} \sim} & \ \textrm{Discrete}(\theta_1,\theta_2,\theta_3)
	\end{aligned}\\
	&\begin{aligned} \label{simulations:probit2}
		\mathllap{X_{i1} | Y_{i1} \sim} & \ \textrm{Bernoulli}(\pi_{X_{i1}}); \ \ \Phi^{-1}(\pi_{X_{i1}}) = \alpha_0 + \alpha_{1j} \mathds{1}[Y_{i1}=j],
	\end{aligned}
\end{align}
for $j\in\{2,3\}$, where $\pi_{X_{i1}} = \Pr[X_{i1} = 1 | Y_{i1}]$.  Here, the Discrete distribution refers to the multinomial distribution with sample size equal to one, and $\Phi^{-1}$ is the inverse cumulative distribution function of the standard normal distribution.   We set $\theta = (\theta_1,\theta_2,\theta_3) = (0.5,0.15,0.35)$ in stratum 1, and $\theta = (0.1,0.45,0.45)$ in stratum 2. This ensures that the joint distributions of $Y_1$ and $X_1$ differ across strata. We set different values for $\alpha_0$, $\alpha_{12}$, and $\alpha_{13}$ to explore how the strength of the relationship between $X_1$ and $Y_1$ affects results.  

For each of the ten simulation runs, we randomly select $n=5000$ observations from the corresponding population using stratified simple random sampling. We sample $n_1 = 1500$ units from stratum 1 and $n_2 = 3500$ units from stratum 2. This disproportionate sampling allocation  ensures that the base weights matter in the estimation of finite population quantities. The survey weights $w_i = N_1/n_1 = 35000/1500 = 23.33$ for all units in stratum 1 and $w_i = N_2/n_2 = 15000/3500 = 4.29$ for all units in stratum 2. 

We introduce item nonresponse in $X_1$ for each of the simulation runs by generating missingness indicators from an AN model. For each $i \in \cal{D}$ in each population, we sample the missingness indicator from a Bernoulli distribution with probability  
\begin{align}
	\phantom{\Phi^{-1}(\Pr[R^x_{i1} = 1 | Y_{i1}, X_{i1}]) =}
	&\begin{aligned} \label{simulations:probit3}
		\mathllap{\Phi^{-1}(\Pr[R^x_{i1} = 1 | Y_{i1}, X_{i1}]) =} & \ \gamma_0 + \gamma_{1j} \mathds{1}[Y_{i1}=j] + \gamma_2 X_{i1},
	\end{aligned}
\end{align}
where $j\in\{2,3\}$.
We set different values for $\gamma_0$, $\gamma_{12}$, $\gamma_{13}$ and $\gamma_2$ to investigate how departures from an ignorable missing mechanism affect the performance of the imputation strategies. All sets result in approximately $30\%$ missing values in $X_1$. 

After making the missing values, we use several approaches to impute the item nonresponse in $X_1$.  For each approach, we use  (\ref{simulations:probit1}) and (\ref{simulations:probit2}) as the models for the survey variables.  We use different methods for specifying and estimating the selection model, in particular for incorporating the weights and auxiliary information.  The approaches include the following. 
\begin{enumerate}
	\item \label{simulations:MAR+Weight}
	MAR+Weight: We incorporate the survey weights by including $w_i$ as a covariate in (\ref{simulations:probit2}). Since there is a one-to-one mapping between weights and strata in our simulation setup, we incorporate $w_i$ by adding an indicator variable $S_i$ for strata, so that we have
	\begin{align}
		\phantom{\Pr[X_{i1} | Y_{i1}] \sim}
		&\begin{aligned} \label{simulations:probit4}
			\mathllap{X_{i1} | Y_{i1} \sim} & \ \textrm{Bernoulli}(\pi_{X_{i1}}); \ \ \Phi^{-1}(\pi_{X_{i1}}) = \alpha_0 + \alpha_{1j} \mathds{1}[Y_{i1}=j] + \alpha_2 \mathds{1}[S_i=2]
		\end{aligned}
	\end{align}
	as the model for $X_1$ instead of (\ref{simulations:probit2}). We exclude the parameter for $\mathds{1}[S_i=1]$ in (\ref{simulations:probit4}) to ensure identifiability. Additionally, since $\gamma_2 X_{i1}$ in (\ref{simulations:probit3}) cannot be identified from the observed data alone, we exclude $\gamma_2 X_{i1}$ in (\ref{simulations:probit3}), so that we have
	\begin{align}
		\phantom{\Phi^{-1}(\Pr[R^x_{i1} = 1 | Y_{i1}, X_{i1}]) =}
		&\begin{aligned} \label{simulations:probit5}
			\mathllap{\Phi^{-1}(\Pr[R^x_{i1} = 1 | Y_{i1}, X_{i1}]) =} & \ \gamma_0 + \gamma_{1j} \mathds{1}[Y_{i1}=j].
		\end{aligned}
	\end{align}
	This is a MAR model for the item nonresponse. This approach represents a default approach analysts might use in this scenario.  It does not use auxiliary information about the margin of $X_1$.
	
	\item \label{simulations:AN+weight}
	AN+Weight: We use (\ref{simulations:probit4}) to incorporate the weights and fit the AN model in (\ref{simulations:probit3}). However, we do so without using any auxiliary information. 
	Although $\gamma_2 X_{i1}$ in (\ref{simulations:probit3}) is not identifiable as previously discussed, the model can be estimated (albeit not accurately) under the Bayesian paradigm because of the prior distribution. This represents a naive application of a nonignorable modeling strategy.
	
	\item \label{simulations:AN+Constraint}
	AN+Constraint: We fit the AN model in (\ref{simulations:probit3}), using the method in Section \ref{framework} to incorporate the auxiliary information and survey design. We incorporate the auxiliary total $T_{X_1}$ and survey weights through the constraint in (\ref{constraint}). We set $V_X$ equal to approximately the theoretical variance of $\hat{T}_X$ without any missing values.  
	
	\item \label{simulations:AN+Constraint+Weight}
	AN+Constraint+Weight: We combine the AN+Weight and AN+Constraint approaches. Specifically, we follow the AN+Constraint method but use (\ref{simulations:probit4}) instead of (\ref{simulations:probit2}) to further control for the weights. 
\end{enumerate}

We use non-informative priors for all parameters.  Specifically, we use the $\textrm{Dirichlet}(1,1,1)$ distribution as the prior distribution for $(\theta_1,\theta_2,\theta_3)$, and a standard multivariate normal distribution as the prior distribution for the set of parameters in each probit model in (\ref{simulations:probit2}) to (\ref{simulations:probit5}).
We fit all models using MCMC sampling. 
We run each MCMC sampler for 10,000 iterations, discarding the first 5,000 as burn-in, resulting in 5,000 posterior samples. We create $L = 50$ multiply imputed datasets, $\textbf{Z} = (\textbf{Z}^{(1)}, \ldots, \textbf{Z}^{(50)})$, from every $100^\textrm{th}$ posterior sample. From each completed dataset $\textbf{Z}^{(l)}$, we compute the design-based estimates of $T_X$, $\alpha_0$, $\alpha_{12}$, and $\alpha_{13}$, along with the corresponding standard errors, using the survey-weighted generalized linear models option in the R package, ``survey''. Although there are differing opinions associated with using survey weights in regression modeling \citep{Pfeffermann1993, Gelman2007}, we use them to ensure all analyses account for the selection effects in the survey design. We also compute estimates of $\gamma_0$, $\gamma_1$ and $\gamma_2$ (which do not depend on the weights by design), along with the corresponding standard errors, from each completed dataset, using the generalized linear models option in the R package, ``stats''.

Within any simulation run, we combine all the estimates across all multiply-imputed datasets using multiple imputation (MI) rules \citep{rubin:1987}. As a brief review of MI, let $q$ be the point estimator of some estimand of interest $Q$ in a completed dataset, and let $u$ be the estimator of its variance.  For $l=1, \dots, L$, let $q_l$ and $u_l$ be the values of $q$ and $u$ in completed dataset $\textbf{Z}^{(l)}$. The MI point estimate of $Q$ is $\bar{q}_L = \sum_{l=1}^L q_l/L$, and the corresponding MI estimate of the variance of $\bar{q}_L$ is given by  $T_L = (1 + 1/L)b_L + \bar{u}_L$, where $b_L = \sum_{l=1}^L (q_l - \bar{q}_L)^2/(L-1)$ and $\bar{u}_L = \sum_{l=1}^L u_l/L$. We write $\bar{q}_L^m$ and $T_L^m$ to represent the values of $\bar{q}_L$ and $T_L$ in the simulation run indexed by $m$, where $m=1, \dots, 10$.

\begin{table}[t!]
	\centering
	\caption{Simulation scenarios presented in Section \ref{simulations}.\label{simscenarios}}
	\begin{tabular}{cccc}
		Scenario & Association $(X_1, Y_1)$ & Departure from ignorable missingness & Margins\\ \hline
		1		&  Strong & Large & Population only\\
		2		&  Weak  & Small  & Population only\\	
		3		&  Strong & Large & Both strata\\	
		4		&  Weak & Small & Both strata\\
	\end{tabular}
\end{table}
We consider eight simulation scenarios resulting from a $2 \times 2 \times 2$ factorial design.  The factors include strong and weak associations among $X_1$ and $Y_1$; large and small departures from ignorable missingness mechanisms; and, margins for $X_1$ known either for the entire population only ($T_X$) or for each of the two strata.  In the interest of space, we report detailed results only for the four scenarios described in Table \ref{simscenarios}.
In each scenario, we report averages of MI estimates across the $10$ runs, including $\sum_{m=1}^{10} \bar{q}_L^m/10$ for the point estimate of each estimand $Q$, and $\sqrt{\sum_{m=1}^{10} T_L^m/10}$ as a measure of the corresponding standard error.  For comparison, we also report results before introduction of missing data, using the average of the ten point estimates and the square root of the average of the variance estimates.

\subsection{Results for scenario 1 and scenario 2} \label{simulations:scenarioone}
In  scenario 1, we set $\alpha_0 = 0.5$, $(\alpha_{12},\alpha_{13}) = (-0.5, -1)$, $\gamma_0 = -0.25$, $(\gamma_{12},\gamma_{13}) = (0.1,0.3)$, and $\gamma_2=-1.1$.  This represents a strong relationship between $Y_1$ and $X_1$, and a nonresponse mechanism that deviates substantially from an ignorable mechanism.  Here, $T_X$ is known only for the entire population and not for the individual strata.

\begin{table}[t!]
	\centering
	\caption{Results for scenario 1: overall auxiliary margin for $X_1$, strong relationship between $Y_1$ and $X_1$ and strong nonignorable nonresponse.}
	\label{simulations:scenarioone:results}
	\subfloat[][HT estimates for $T_X$ under each method, the corresponding standard errors, and acceptance ratios.  ``Population'' is the value of $T_X$ in the population of $N=50000$ individuals. ``No Missing Data'' is the weighted estimate based on the sampled data before introducing item nonresponse. For AN+Constraint and AN+Constraint+Weight, the estimated Monte Carlo standard errors of $\sum_{m=1}^{10} \bar{q}_L^m/10$ are less than 150, ruling out chance error as explanation for the improved performance of these two models over AN+Weight and MAR+Weight. 
	\vspace{10pt}]{\renewcommand\arraystretch{1}
		\begin{tabular}[c]{L{3cm}R{1cm}R{1cm}R{1cm}R{1.5cm}}
			\toprule
			& \multicolumn{2}{c}{$T_X$} & \multicolumn{2}{c}{Acceptance Ratio} \\ \cmidrule(lr){2-3} \cmidrule(lr){4-5}
			Method & Mean & SE & Mean & Range \\
			\midrule
			Population  &  25026 &  --- &  --- &  --- \\ 
			Mo Missing Data & 25275 & 582  &  --- &  ---\\ 
			MAR+Weight  & 30579 & 670 & --- &  ---\\ 
			AN+Weight  & 28222 & 2789 & --- &  ---\\ 
			AN+Constraint  & 24993 & 741 & .82 & [.79, .84] \\ 
			AN+Constraint+Weight  & 25019 & 718 & .83 & [.80, .86] \\ 
			\bottomrule
	\end{tabular}}
	\qquad \qquad  \qquad 
	\subfloat[][Survey-weighted estimates of $\alpha_0$, $\alpha_{12}$, $\alpha_{13}$, $\gamma_0$, $\gamma_{12}$, $\gamma_{13}$ and $\gamma_2$, along with the corresponding standard errors. ``MAR+W'' is MAR+Weight, ``AN+W'' is AN+Weights, ``AN+C'' is AN+Constraint, and ''AN+C+W'' is ''AN+Constraint+Weight''.  Standard errors of the averaged point estimates are small enough to rule out chance error as explanations for the improved performance of AN+C and AN+C+W over MAR+W and AN+W.]{\renewcommand\arraystretch{1}
		\begin{tabular}[c]{C{0.6cm}R{0.85cm}R{0.85cm}R{0.6cm}R{0.85cm}R{0.6cm}R{0.85cm}R{0.6cm}R{0.85cm}R{0.6cm}}
			\toprule
			& &
			\multicolumn{2}{c}{MAR+W} &
			\multicolumn{2}{c}{AN+W} &
			\multicolumn{2}{c}{AN+C}  &
			\multicolumn{2}{c}{AN+C+W} 
			\\ \cmidrule(lr){3-4} \cmidrule(lr){5-6} \cmidrule(lr){7-8} \cmidrule(lr){9-10}  
			Par. & Truth & Mean & SE & Mean & SE & Mean & SE & Mean & SE \\
			\midrule
			$\alpha_{0}$  & .50 &
			.74 & .05 &
			.63 & .13 &
			.49 & .05 &
			.49 & .05 \\
			$\alpha_{12}$  & -.50 &
			-.45 & .07 &
			-.47 & .07 &
			-.49 & .07 &
			-.49 & .06 \\
			$\alpha_{13}$  & -1.00 &
			-.88 & .07 &
			-.92 & .10 &
			-.98 & .06 &
			-.98 & .06 \\
			\midrule 
			$\gamma_{0}$  & -.25 &
			-.88 & .04 &
			-.63 & .35 &
			-.22 & .07 &
			-.23 & .07 \\
			$\gamma_{12}$  & .10 &
			.29 & .05 &
			.21 & .11 &
			.10 & .06 &
			.10 & .06 \\
			$\gamma_{13}$  & .30 &
			.63 & .05  &
			.48 & .17 &
			.27 & .07 &
			.27 & .07 \\
			$\gamma_{2}$  & -1.10 &
			--- &  --- &
			-.48 & .57 &
			-1.15 & .14 &
			-1.15 & .13 \\
			\bottomrule
	\end{tabular}}
\end{table}
For each method, Table \ref{simulations:scenarioone:results}(a) displays the average of the ten HT estimates for $T_X$ and the square root of the average of the variances of these estimates in scenario 1. AN+Constraint and AN+Constraint+Weight offer the most accurate estimates, whereas AN+Weight and MAR+Weight offer the least accurate estimates. 
Controlling for the weights in the model for $X_1$ as in the AN+Constraint+Weight method apparently decreases the standard error in comparison to AN+Constraint.  It also increases the acceptance ratios in the MCMC samplers. The standard error associated with AN+Weight is much higher than all other methods. This is due primarily to the weak identification issues associated with using the AN model without any auxiliary information, resulting in greater uncertainty from the nonresponse mechanism.

Table \ref{simulations:scenarioone:results}(b) also shows survey-weighted estimates of $\alpha_0$, $\alpha_{12}$, $\alpha_{13}$, $\gamma_0$, $\gamma_{12}$, $\gamma_{13}$ and $\gamma_2$, along with the corresponding standard errors, again combined across all ten simulation runs. Here, both AN+Constraint and AN+Constraint+Weight give nearly identical results and closely estimate the true parameter estimates.
The AN+Constraint and AN+Constraint+Weight approaches outperform the other choices in this scenario. AN+Weight and MAR+Weight again give the least accurate results.

In scenario 2, we weaken both the relationship between the variables of interest and the nonignorable nonresponse. We set $\alpha_0 = 0.15$ and $(\alpha_{12},\alpha_{13}) = (-0.45, -0.15)$ to reflect a weak relationship between $Y_1$ and $X_1$, and we set $\gamma_0 = -1$, $(\gamma_{12},\gamma_{13}) = (-0.6,1.4)$ and $\gamma_2=-0.2$ to reflect a small departure from an ignorable nonresponse mechanism.  $T_X$ is known only for the entire population.

\begin{table}[t!]
	\centering
	\caption{Results for scenario 2: overall auxiliary margin for $X_1$, weak relationship between $Y_1$ and $X_1$ and weak nonignorable nonresponse.}
	\label{simulations:scenariotwo:results}
	\subfloat[][HT estimates for $T_X$ under each method, the corresponding standard errors, and acceptance ratios.  ``Population'' is the value of $T_X$ in the population of $N=50000$ individuals. ``No Missing Data'' is the weighted estimate based on the sampled data before introducing item nonresponse. 
	For AN+Constraint and AN+Constraint+Weight, the estimated Monte Carlo standard errors of $\sum_{m=1}^{10} \bar{q}_L^m/10$ are less than 150, ruling out chance error as explanation for the improved performance of these two models over AN+Weight and MAR+Weight.
	\vspace{10pt}]{\renewcommand\arraystretch{1}
		\begin{tabular}[c]{L{3cm}R{1cm}R{1cm}R{1cm}R{1.5cm}}
			\toprule
			& \multicolumn{2}{c}{$T_X$} & \multicolumn{2}{c}{Acceptance Ratio} \\ \cmidrule(lr){2-3} \cmidrule(lr){4-5}
			Method & Mean & SE & Mean & Range \\
			\midrule
			Population  & 24677  &  --- &  --- &  --- \\ 
			Mo Missing Data & 24742 &  570 &  --- &  ---\\ 
			MAR+Weight  & 26098 & 662 & --- &  ---\\ 
			AN+Weight  & 23705 & 2519 & --- &  ---\\ 
			AN+Constraint  & 24666 & 698 &  .79 & [ .77,  .81] \\ 
			AN+Constraint+Weight  & 24653 & 705 &  .79 & [ .77,  .81] \\ 
			\bottomrule
	\end{tabular}}
	\qquad 
	\subfloat[][Survey-weighted estimates of $\alpha_0$, $\alpha_{12}$, $\alpha_{13}$, $\gamma_0$, $\gamma_{12}$, $\gamma_{13}$ and $\gamma_2$, along with the corresponding standard errors. ``MAR+W'' is MAR+Weight, ``AN+W'' is AN+Weights, ``AN+C'' is AN+Constraint, and ''AN+C+W'' is ''AN+Constraint+Weight''. Standard errors of the averaged point estimates are small enough to rule out chance error as explanations for the improved performance of AN+C and AN+C+W over MAR+W and AN+W.]{\renewcommand\arraystretch{1} 
		\begin{tabular}[c]{C{0.6cm}R{0.85cm}R{0.85cm}R{0.6cm}R{0.85cm}R{0.6cm}R{0.85cm}R{0.6cm}R{0.85cm}R{0.6cm}}
			\toprule
			& &
			\multicolumn{2}{c}{MAR+W}  &
			\multicolumn{2}{c}{AN+W} &
			\multicolumn{2}{c}{AN+C}  &
			\multicolumn{2}{c}{AN+C+W} \\
			\cmidrule(lr){3-4} \cmidrule(lr){5-6} \cmidrule(lr){7-8} \cmidrule(lr){9-10}  
			Par. & Truth & Mean & SE & Mean & SE & Mean & SE & Mean & SE \\
			\midrule
			$\alpha_{0}$  &  .15 &
			.19 &   .05 &
			.12 &  .08 &
			.15 &  .05 &
			.15 &  .05 \\
			$\alpha_{12}$  & -.45 &
			-.48 &  .06 &
			-.43 &  .08 &
			-.45 &  .06 &
			-.45 &  .06 \\
			$\alpha_{13}$  & -.15 &
			-.04 &  .07 &
			-.23 &  .21 &
			-.15 &  .07 &
			-.16 &  .07 \\
			\midrule 
			$\gamma_{0}$  & -1.00 &
			-1.12 &   .05 &
			-.97 &  .21 &
			-1.00 &  .06 &
			-1.00 &  .06 \\
			$\gamma_{12}$ & -.60 &
			-.57 &   .07 &
			-.64 &  .10 &
			-.61 &  .07 &
			-.61 &  .07 \\
			$\gamma_{13}$  & 1.40 &
			1.42 &   .06 &
			1.41 &  .06 &
			1.42 &  .06 &
			1.42 &  .06 \\
			$\gamma_{2}$  & -.20 &
			--- &  --- &
			-.44 &  .47 &
			-.23 &  .08 &
			-.23 &  .08 \\
			\bottomrule
	\end{tabular}}
\end{table}

Tables \ref{simulations:scenariotwo:results}(a) and \ref{simulations:scenariotwo:results}(b) present results of 10 simulation runs of scenario 2. Once again, the AN+Constraint and AN+Constraint+Weight outperform the other methods. AN+Constraint has a slightly smaller standard error for $T_X$ in scenario 2 than AN+Constraint+Weight.  
Also, MAR+Weight performs much better in scenario 2 than in scenario 1. In the presence of a weakly nonignorable nonresponse mechanism, there appears to be little degradation when using a MAR model. In addition, whatever degradation or bias that should have been attributed to the survey design appears to be taken care of by including the strata indicator in the model for $X_1$. AN+Weight performs worse than the other three methods. Unlike before, AN+Weight actually underestimates rather than overestimates $T_X$ in this scenario. Overall, the range of acceptance ratios has decreased slightly from the previous scenario. 

We note that we find similar overall conclusions in the two other scenarios where we know the margin of $T_X$ only for the whole population.

\subsection{Results for scenario 3 and scenario 4} \label{simulations:scenariothree}

We next investigate the performance of the approaches when we know the auxiliary margin of $X_1$ in each stratum.
In this case, it is possible to implement the constraint in (\ref{constraint}) for each stratum. 
For each stratum $s \in \{1, 2\}$, we require that
\begin{equation} \label{constraint:bystrata}
	\sum_{\substack{S_i = s; \\i \in \cal{D}}} w_iX_{i1}^\star = \dfrac{N_s}{n_s} \sum_{\substack{S_i = s; \\i \in \cal{D}}} X_{i1}^\star \sim N(T_X^{(s)}, V_X^{(s)}),
\end{equation}
where $T_X^{(s)}$ is the auxiliary total of $X_1^{pop}$ for stratum $s$, and $V_X^{(s)}$ is the corresponding variance associated with it. 
For the AN+Constraint and AN+Constraint+Weight methods, we implement this constraint by applying the Metropolis steps S1 to S3 in Section \ref{framework} within each stratum. 

\begin{table}[t!]
	\centering
	
	\caption{Results for scenario 3: auxiliary margin for $X_1$ within each stratum, strong relationship between $Y_1$ and $X_1$ and strong nonignorable nonresponse.}
	\label{simulations:scenariothree:results}
	\subfloat[][HT estimates for $T_X$ under each method, the corresponding standard errors, and acceptance ratios by strata.  ``Population'' is the value of $T_X$ in the population of $N=50000$ individuals. ``No Missing Data'' is the weighted estimate based on the sampled data before introducing item nonresponse. 
	For AN+Constraint and AN+Constraint+Weight, the estimated Monte Carlo standard errors of $\sum_{m=1}^{10} \bar{q}_L^m/10$ are less than 250, ruling out chance error as explanation for the improved performance of these two models over AN+Weight and MAR+Weight.
	\vspace{10pt}]{\renewcommand\arraystretch{1}
		\begin{tabular}[c]{L{3cm}R{1cm}R{1cm}R{1cm}R{1.5cm}R{1cm}R{1.5cm}}
			\toprule
			& \multicolumn{2}{c}{$T_X$} & \multicolumn{4}{c}{Acceptance Ratio} \\ \cmidrule(lr){2-3} \cmidrule(lr){4-7}
			& \multicolumn{2}{c}{} & \multicolumn{2}{c}{\underline{Stratum 1}} & \multicolumn{2}{c}{\underline{Stratum 2}}\\
			Method & Mean & SE & Mean & Range & Mean & Range \\
			\midrule
			Population  & 24994 &  --- &  --- &  ---  &  --- &  --- \\ 
			Mo Missing Data & 25043 &  580 &  --- &  --- &  --- &  --- \\ 
			MAR+Weight  & 30447 & 668 & --- &  --- &  --- &  --- \\ 
			AN+Weight  & 28488 & 3034 & --- &  --- &  --- &  --- \\ 
			AN+Constraint  & 25062 & 665 &  .81 & [ .66,  .91] &  .80 & [ .74,  .83] \\ 
			AN+Constraint+Weight  & 25070 & 667 &  .80 & [ .61,  .90] &  .79 & [ .74,  .84] \\ 
			\bottomrule
	\end{tabular}}
	\qquad 
	\subfloat[][Survey-weighted estimates of $\alpha_0$, $\alpha_{12}$, $\alpha_{13}$, $\gamma_0$, $\gamma_{12}$, $\gamma_{13}$ and $\gamma_2$, along with the corresponding standard errors. ``MAR+W'' is MAR+Weight, ``AN+W'' is AN+Weights, ``AN+C'' is AN+Constraint, and ''AN+C+W'' is ''AN+Constraint+Weight''. Standard errors of the averaged point estimates are small enough to rule out chance error as explanations for the improved performance of AN+C and AN+C+W over MAR+W and AN+W.]{\renewcommand\arraystretch{1} 
		\begin{tabular}[c]{C{0.6cm}R{0.85cm}R{0.85cm}R{0.6cm}R{0.85cm}R{0.6cm}R{0.85cm}R{0.6cm}R{0.85cm}R{0.6cm}}
			\toprule
			& &
			\multicolumn{2}{c}{MAR+W}  &
			\multicolumn{2}{c}{AN+W} &
			\multicolumn{2}{c}{AN+C}  &
			\multicolumn{2}{c}{AN+C+W} \\
			\cmidrule(lr){3-4} \cmidrule(lr){5-6} \cmidrule(lr){7-8} \cmidrule(lr){9-10} 
			Par. & Truth & Mean & SE & Mean & SE & Mean & SE & Mean & SE \\
			\midrule
			$\alpha_{0}$  &  .50 &
			.74 &  .05 &
			.64 &  .13 &
			.50 &  .05 &
			.50 &  .05 \\
			$\alpha_{12}$ & -.50 &
			-.45 &  .07 &
			-.46 &  .08 &
			-.50 &  .07 &
			-.50 &  .07 \\
			$\alpha_{13}$ & -1.00 &
			-.89 &  .07 &
			-.90 &  .12 &
			-1.00 &  .06 &
			-1.00 &  .07 \\
			\midrule 
			$\gamma_{0}$  & -.25 &
			-.89 &  .04 &
			-.73 &  .44 &
			-.27 &  .06 &
			-.27 &  .06 \\
			$\gamma_{12}$ &  .10 &
			.30 &  .05 &
			.22 &  .11 &
			.12 &  .06 &
			.12 &  .06 \\
			$\gamma_{13}$  &  .30 &
			.65 &  .05 &
			.52 &  .19 &
			.31 &  .06 &
			.31 &  .06 \\
			$\gamma_{2}$  & -1.10 &
			--- & --- &
			-.41 &  .69 &
			-1.08 &  .09 &
			-1.08 &  .09 \\
			\bottomrule
	\end{tabular}}
\end{table}
We first set the parameters as in Section \ref{simulations:scenarioone} to reflect a strong relationship between $Y_1$ and $X_1$, and strong nonignorable nonresponse mechanism.
Table \ref{simulations:scenariothree:results}(a) shows the average HT estimates for $T_X$, the standard error under each method and the acceptance ratios by strata. Table \ref{simulations:scenariothree:results}(b) shows survey-weighted estimates of $\alpha_0$, $\alpha_{12}$, $\alpha_{13}$, $\gamma_0$, $\gamma_{12}$, $\gamma_{13}$ and $\gamma_2$, and the corresponding standard errors. The overall conclusions are qualitatively similar to those in Section \ref{simulations:scenarioone}. Incorporating the auxiliary margin by strata in AN+Constraint and AN+Constraint+Weight reduces the standard errors. 
AN+Weight and MAR+Weight again yield the least accurate results. The range of acceptance ratios are much wider suggesting that there is a smaller set of combinations of imputed values that fulfill the constraints within each stratum, than with the combined constraint.

\begin{table}[t!]
	\centering
	
	\caption{Results for scenario 4: auxiliary margin for $X_1$ within each stratum, weak relationship between $Y_1$ and $X_1$ and weak nonignorable nonresponse.}
	\label{simulations:scenariofour:results}
	\subfloat[][HT estimates for $T_X$ under each method, the corresponding standard errors, and acceptance ratios by strata.  ``Population'' is the value of $T_X$ in the population of $N=50000$ individuals. ``No Missing Data'' is the weighted estimate based on the sampled data before introducing item nonresponse. For AN+Constraint and AN+Constraint+Weight, the estimated Monte Carlo standard errors of $\sum_{m=1}^{10} \bar{q}_L^m/10$ are less than 350.
	\vspace{10pt}]{\renewcommand\arraystretch{1}
		\begin{tabular}[c]{L{3cm}R{1cm}R{1cm}R{1cm}R{1.5cm}R{1cm}R{1.5cm}}
			\toprule
			& \multicolumn{2}{c}{$T_X$} & \multicolumn{4}{c}{Acceptance Ratio} \\ \cmidrule(lr){2-3} \cmidrule(lr){4-7}
			& \multicolumn{2}{c}{} & \multicolumn{2}{c}{\underline{Stratum 1}} & \multicolumn{2}{c}{\underline{Stratum 2}}\\
			Method & Mean & SE & Mean & Range & Mean & Range \\
			\midrule
			Population  &  24724 &  --- &  --- &  ---  &  --- &  --- \\ 
			Mo Missing Data & 24613 & 569  &  --- &  --- &  --- &  --- \\ 
			MAR+Weight  & 25969 & 669 & --- &  --- &  --- &  --- \\ 
			AN+Weight  & 23551 & 3038 & --- &  --- &  --- &  --- \\ 
			AN+Constraint  & 24710 & 651 &  .79 & [ .60,  .87] &  .76 & [ .68,  .79] \\ 
			AN+Constraint+Weight  & 24689 & 672 &  .77 & [ .59,  .86] &  .75 & [ .66,  .78] \\ 
			\bottomrule
	\end{tabular}}
	\qquad 
	\subfloat[][Survey-weighted estimates of $\alpha_0$, $\alpha_{12}$, $\alpha_{13}$, $\gamma_0$, $\gamma_{12}$, $\gamma_{13}$ and $\gamma_2$, along with the corresponding standard errors. ``MAR+W'' is MAR+Weight, ``AN+W'' is AN+Weights, ``AN+C'' is AN+Constraint, and ''AN+C+W'' is ''AN+Constraint+Weight''. 
	]{\renewcommand\arraystretch{1} 
		\begin{tabular}[c]{C{0.6cm}R{0.85cm}R{0.85cm}R{0.6cm}R{0.85cm}R{0.6cm}R{0.85cm}R{0.6cm}R{0.85cm}R{0.6cm}}
			\toprule
			& &
			\multicolumn{2}{c}{MAR+W}  &
			\multicolumn{2}{c}{AN+W} &
			\multicolumn{2}{c}{AN+C}  &
			\multicolumn{2}{c}{AN+C+W} \\
			\cmidrule(lr){3-4} \cmidrule(lr){5-6} \cmidrule(lr){7-8} \cmidrule(lr){9-10}  
			Par. & Truth & Mean & SE & Mean & SE & Mean & SE & Mean & SE \\
			\midrule
			$\alpha_{0}$  &  .15 &
			.18 &  .05  &
			.12 &  .09 &
			.15 &  .05 &
			.15 &  .05 \\
			$\alpha_{12}$  & -.45 &
			-.48 &  .06  &
			-.44 &  .08 &
			-.46 &  .06 &
			-.46 &  .06 \\
			$\alpha_{13}$  & -.15 &
			-.05 &   .07 &
			-.24 &  .25 &
			-.14 &  .07 &
			-.14 &  .07 \\
			\midrule 
			$\gamma_{0}$  & -1.00 &
			-1.09 &  .05  &
			-.97 &  .26 &
			-.98 &  .06 &
			-.98 &  .06 \\
			$\gamma_{12}$ & -.60 &
			-.60 &  .07  &
			-.68 &  .12 &
			-.64 &  .07 &
			-.64 &  .07 \\
			$\gamma_{13}$  & 1.40 &
			1.38 &   .05 &
			1.37 &  .06 &
			1.38 &  .06 &
			1.38 &  .06 \\
			$\gamma_{2}$  & -.20 &
			--- &  --- &
			.50 &  .62 &
			-.21 &  .07 &
			-.21 &  .07 \\
			\bottomrule
	\end{tabular}}
\end{table}
We also set the parameters as in Section \ref{simulations:scenarioone} to reflect a weak relationship between $Y_1$ and $X_1$, and a weakly nonignorable nonresponse mechanism. Tables \ref{simulations:scenariofour:results}(a) and \ref{simulations:scenariofour:results}(b) display the results.  The conclusions are qualitatively similar to those in previous simulations. The primary difference is that implementing the constraint by strata reduces the standard errors for AN+Constraint and AN+Constraint+Weight.  

Results for the remaining two scenarios with known population totals per stratum are qualitatively similar to those presented here.

\section{Discussion} \label{discussion}

The results  suggest that the approach in Section \ref{framework} can allow survey analysts to incorporate survey weights and auxiliary information when imputing nonresponse in complex surveys. In particular, AN+Constraint and AN+Constraint+Weight appear to outperform the default option of controlling for the weights in the joint model for the variables in $\cal{D}$.  The MAR+Weight approach offers good results when the nonresponse mechanism is only weakly nonignorable; we expect that this method should perform even better for fully ignorable nonresponse mechanisms. However, the results based on AN+Constraint and AN+Constraint+Weight are the most consistently best across the different scenarios. Of course, these results are based on a limited set of simulation scenarios, and the methods could perform differently in other scenarios.

Opportunities for extensions of this approach exist as future research topics. First, future work could explore extensions of the approach to other sampling designs, in particular when weights have many unique values. Preliminary simulations, not shown here, suggest that our approach also can work well for many valued, unequal weights. However, generating plausible imputations that satisfy the constraint can be challenging whenever the set of combinations of imputed values that result in completed datasets that satisfy the constraint is small compared to the set of all possible combinations. When this is the case, we have found that one needs efficient samplers for generating proposals for the imputations.  Finding general strategies for such proposals is an important topic for future work.

Second, as we suggested in Section \ref{framework}, there are opportunities to investigate different approaches to specifying the constraints involving $\hat{T}_X$, in particular how to set the variance $V_X$.  In the simulations we used the theoretical design-based variance, estimated via resampling from the true generative process, but this would need to be approximated in practice.  Future research could examine the effectiveness of using the types of approximations described in Section \ref{framework}. Additionally, one could investigate how different values of $V_X$ affect the performance of the methodology. For example, using very small $V_X$ could lead to a more efficient estimation of $T_X$; however, forcing the completed datasets to match very closely on $T_X$ could affect the relationships among $X$ and $Y$ in the completed data in unpredictable ways.  It also could lead to a less efficient MCMC sampler, since the set of imputations consistent with (\ref{constraint}) would be smaller. 

Third, future research could adapt this approach to other model specifications.  For example,  
one could extend the approach to nonparametric models and semi-parametric models like those in \citet{KimYu2011, MorikawaEtAl2017}.

Fourth, one could extend the framework to handle imputation for unit nonresponse as well.  In particular, we conjecture that analysts can follow the framework developed by \citet[Chapter 4]{Akande2019}, who extends the AN model to unit nonresponse as well as item nonresponse in more than one variable in simple random samples.  We expect that analysts can add the probabilistic constraint on the completed-data totals on top of the models in \citet[Chapter 4]{Akande2019}. 
We note that this requires survey weights for the unit nonrespondents, which often are not available.

Finally, we work with base weights instead of more complex ``adjusted'' weights, which are often inflated to adjust for nonresponse or poststratification. Since we take a model-based approach to handling survey nonresponse, there is no obvious justification for using adjusted weights that already account for the nonresponse. In fact, using such adjusted weights assumes that the weights are fixed, which is not often true as pointed out by \citet{Fienberg2010}. 
Since agencies often release those adjusted weights in practice, instead of the base weights, future work would explore the extension of our approach to adjusted weights as well.

\begin{acknowledgement}
	This research was supported by grants from the National Science Foundation (SES-1131897 and SES-1733835).
\end{acknowledgement}

\printbibliography

\end{document}